# On the Naming of Methods: A Survey of Professional Developers

Reem S. AlSuhaibani
*Computer Science*
Kent State University
Ohio, USA
ralsuhai@kent.edu

Christian D. Newman
*Software Engineering*
Rochester Institute of Technology
New York, USA
cnewman@se.rit.edu

Michael J. Decker
*Computer Science*
Bowling Green State University
Ohio, USA
mdecke@bgsu.edu

Michael L. Collard
*Computer Science*
The University of Akron
Ohio, USA
collard@uakron.edu

Jonathan I. Maletic
*Computer Science*
Kent State University
Ohio, USA
jmaletic@kent.edu

*Abstract*—This paper describes the results of a large (+1100 responses) survey of professional software developers concerning standards for naming source code methods. The various standards for source code method names are derived from and supported in the software engineering literature. The goal of the survey is to determine if there is a general consensus among developers that the standards are accepted and used in practice. Additionally, the paper examines factors such as years of experience and programming language knowledge in the context of survey responses. The survey results show that participants very much agree about the importance of various standards and how they apply to names. Additionally, the survey shows that years of experience and the programming language the participants use has almost no effect on their responses.

## I. INTRODUCTION

*"If you have a good name for a method, you do not need to look at the body"* - Fowler et al. [1]

The naming of source code identifiers (e.g., variables, types, methods, functions, classes, etc.) is a critical issue for software engineering. It is discussed from day one in introductory programming courses, and it is argued about daily during code reviews of commercial and open-source projects alike. Software engineers make decisions about names constantly. Naming impacts the readability and comprehension [2-6] of software. Good names reduce the cost of software maintenance [7,8]. Careful selection of names can convey the high-level meaning of the task to the developer [9].

While there are many implied and documented standards for naming, there is no broad understanding of how these standards are used or accepted by developers. Standards include, but are not limited to, the allowable words in names, the grammatical structure of the name, the number of words in a name, and how multi-word names are composed. The goal of this work is to determine if the standards documented in software engineering literature reflect actual practice and align with developer opinion.

Here, we focus specifically on the names given to methods in object-oriented software systems. However, much of this also applies to (free) functions in non-object-oriented systems (or parts). We focus on methods for several reasons. First, we are interested in method naming in the context of automatic method summarization and documentation. Furthermore, different programming language constructs have their own naming standards. That is, local variable names are named differently than method names, which are named differently than class names [10,11]. Of these, prior work has found that function names have the largest number of unique name variants when analyzed at the level of part-of-speech sequences [12]; implying that method names are more complex on average than other types of identifiers. We think that prioritizing a focused survey on method names will best serve the research community and possibly motivate future research on other types of identifiers. Lastly, focusing on just method names makes the survey short enough to be completed by a larger number of participants.

Our motivations for undertaking the articulation of method naming standards is to construct automated tools to assess the quality of method names. These tools will be clearly useful for developers daily naming, renaming, and performing code review activities. Following standards for naming methods has a large impact on the quality of software [13, 14]. Furthermore, this quality information can be leverage in current research for automated code summarization [15-17], part-of-speech tagging [18,19], topic modeling [20], feature location [21, 22], concept location [23], code search tools [24-26] identifier splitting [27, 28] and other natural language analysis tools such as [29].

There are a large number of naming practices which the research community has derived and suggested to developers. This advice, put together, forms a standard for naming methods. Given this standard, derived from the literature, we feel it is necessary to assess its quality and practicality. As such, we developed a survey to rate each aspect of the standard. Additionally, we feel that professional developers must be the vast majority of the participant pool in order to have valid and meaningful results. Surveys are an appropriate empirical strategy to gather data from a large population [30]. The participants in this survey are a representative sample of software engineering developers. The goal is to find statistical outcomes about how much agreement and disagreement participants have.

Our hypothesis is that the results of the survey will convey a naming standard for methods that is used in practice and widely accepted by developers. The work aims to address the following research questions:

RQ1: To what extent do software professionals support the method naming standards conveyed in the survey? That is, do developers agree with each part of the standard. Is the standard complete?

RQ2: Do years of programming experience impact how professionals respond to the survey questions? That is, do senior developers have the same responses as junior developers. Or is there some change in attitude that comes with experience?

RQ3: Does the programming language used impact how professionals respond to the survey questions? That is, does the programming language a developer normally uses impact on how they name methods?

RQ4: What are professionals' perceptions of each part of the method naming standard in the survey? What are the preferences and barriers concerning each part of the naming standard?

The contributions of this work are as follows:
- The results of the survey show that professional developers are very much in agreement with the method naming standards. This validates the standard to a great extent.
- The results also show that the standard is complete to a great extent. However, the results uncover a couple of special cases that need to be articulated.
- We also find that years of experience and programming language knowledge have no impact on how the participants responded to the survey.

The paper is organized as follows. The next section describes method naming standards and the related literature on coding standards. Section III describes the design of the survey Section IV describes and discusses the results. Threats to validity are given in Section V, and Section VI gives conclusions.

## II. Standards for Method Names

We now define the method naming standard in detail. The standard is developed by examining the related software engineering literature and exiting coding standards. There exists a wealth of standards and discussions on the topic.

Leading industrial professionals Martin[31], Beck [32], and McConnell [33] insist on the importance of identifier naming and discus multiple tips for developers. Pavlutin [34] discusses some practical function naming conventions and motivates its importance to code readability. Devopedia [35] also discusses the advantages of naming conventions and provides an overview of common naming conventions used in programming. Tan [36] explains general naming rules he established over the years, and Piater [37] also provides guidelines on coding standards for maintainable code.

While not specific to only method names Relf [38] investigates 21 identifier naming style guidelines that focus on the typography and length of identifiers with some real-world examples from Java and Ada to illustrate compliance and non-compliance. He also investigates the attitudes of industry software engineers toward the acceptance of these guidelines. Hilton [39], reviews these guidelines in his blog and provides some perceptions.

In another general look at identifier naming styles, Butler et al. [14] investigate 12 style guidelines. Butler [40] also studies mining Java class identifier naming conventions. He investigates the structure of Java class names and identifies common naming patterns. Butler et al. [10] conduct a survey of the forms of Java reference names and then use the study outcome to investigate naming convention adherence in Java references [41]. Arnaoudova et al. [42, 43] defined source code Linguistic Antipatterns (LAs) that discuss poor practices in naming and choice of identifiers and defined a catalog of 17 types of LAs related to inconsistencies.

From this work, we identified 10 core standards that apply to method names. These standards cover the types of words used, the grammatical structure, and the overall length of the method name. Together, they form a standard derived through research. The standard represents a set of guidelines or heuristics that can be used to assess the quality of a method name. That is, names that adhere to the standard are, based on prior research and naming guidelines, most likely reasonable. Names that fall outside one or more aspects of the standard need to be reviewed. Each is now defined individually, along with supporting literature and examples.

### A. Naming Style

This standard involves using a consistent naming style of methods for the entire application. There are two main naming styles regularly used by developers for names with two or more words. The two popular styles are camelCase and under_score. There are variants and/or other names for both. The first is camelCase, where the first letter of each word is capitalized, starting with the second word. A variant of this is PascalCase, where the first letter of every word is capitalized. The next is under_score, which uses an underscore between words. Another name for this style is *snake case*. A variant of this is *kebab-case* (used in Lisp and Forth), which uses a dash instead of an underscore.

While there is much discussion on which is the better style, camelCase or under_score, an in-depth study [44] shows very little difference between the two in terms of increased or decreased cognitive load during comprehension. Any differences appear to be mitigated by training in a particular style. However, the study does show that camelCase appears to have a slight advantage for the comprehension of shorter identifier names. Binkley et al. [45] also discussed naming styles, i.e., camelCase and under_score, in recognizing source code identifier names for better readability and comprehension. Maletic and Sharif conducted an eye-tracking study on camelCase and under_score identifier styles [46]. The bottom line is that a consistent naming style should be defined and used within a given application.

Some examples that follow one of these standards include: `getFullName()`, `getScriptState()`, `call_with_default()`, `garbage_collection()` and `check_static_allocation_size()`. Examples that do not follow the standard includes `getfullName()`, `getscriptstate()`.

### B. Grammatical Structure

Method names with multiple words need to have a grammatically correct sentence structure (see Fig. 1). For example, in method names without a preposition, there is a sequence of words to the left of a head-noun; one of which is a verb (typically the first word) [12, 47, 48]. That is, the name should be a grammatically meaningful verb phrase; the words from left to right should modify a head-noun somehow. There is research that discusses the grammatical structure of identifiers and common grammar patterns and how words are related. Caprile & Tonella [49], focus on analyzing the grammar of function identifiers. Among other things, they find several patterns for function identifiers and create a formal grammar of each pattern. While not all patterns are verb phrases, many contain verb phrases. In addition, there is a finite set of common, diverse grammatical structures that convey different types of actions such as conversion, predicate logic, etc. Thus, they find that there is a set of specific grammatical structures to method names.

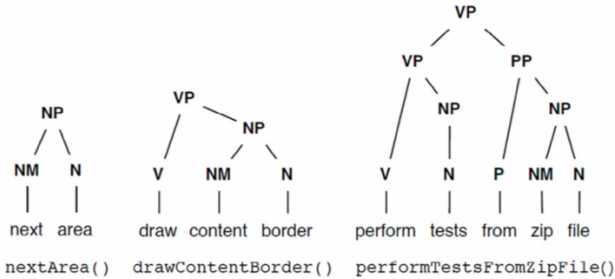

Fig. 1. An example of a noun phrase, a verb phrase, and a verb phrase with a prepositional phrase for three method name identifiers.

Deisenbock and Pizka [50] discuss naming structure. In part, they find that identifier names tend to be compound-words, where the words to the left in the identifier specialize the meaning of the word that comes to the right, referred to as the head. They believe that these regularities reveal additional meaning about the code. Newman et al. [12] explore the grammatical structure of identifiers by manually annotating a sample of 1,335 identifier names separated by where they appear in the code (i.e., function name, class name, etc.). They then discuss common naming structures for different types of identifiers, based on sequences of part-of-speech tags they refer to as grammar patterns. Among their findings, they highlight the ubiquity of noun-adjuncts (nouns used as adjectives) and their usage in noun and verb phrases to modify the meaning of their corresponding head-noun. They also highlight several distinct grammar patterns and how the implementation of a method influences the interpretation and construction of its name.

An example of the standard is registerManagedResource(). The grammatical structure of this name is a correct sentence that starts with a verb followed by an adjective and ends with a noun. An example name that does not follow the standard is managedResourceRegister(). The grammatical structure of this name is not a correct sentence as it starts with an adjective followed by a noun and ends with a verb/noun.

*C. Verb Phrase*

This standard requires that a given method name not only be grammatically correct (per the previous standard) but also contain a verb or be a verb phrase (see Fig. 1). Liblit et al. [51] assert that methods are actions and mathematical functions that passively compute a result, and therefore, names for such methods should be a verb phrase. Abebe et al. [52], assert that method identifiers should start with a verb, as they believe that an identifiers' grammatical structure should be appropriate for the specific type of entity it represents; thus, method names should contain an action. According to Etzkorn et al. [53], a study of identifiers in C++ code shows that member function names tend to contain verbs and can often be put in sentence form, containing a subject and a verb. Fry et al. [54] and Shepherd et al. [55, 56] studied verb direct object pairs to locate action-oriented concerns in method names. Shepherd et al. [57] focused on creating techniques for extracting verbs from method names, which they used in several natural language processing based tools. The Java Language Specification recommends that method names should be verbs or verb phrases [58].

Some examples that follow this standard include manage_caching_sizes(), computeProductBlockingSizes() and get_cached_node(). An example that does not follow the standard is x_cached_node(), where 'cached' is not a past-tense verb, but an adjective to modify the meaning of the head-noun 'node'. It is not uncommon for some verbs to be used as adjectives in identifiers [12, 59, 60].

*D. Dictionary Terms*

This standard requires that the words used in the method name be actual dictionary terms. That is, the words should be meaningful natural language terms rather than non-dictionary terms; sequences of characters that are not in a standard English or domain dictionary (e.g., using the number '2' to mean 'to'). Deisenbock & Pizka [47], argue that poor naming in one part of the code spoils comprehension in numerous other parts of the code. Thus, the use of terms that are not defined in a dictionary makes an otherwise high-quality identifier hard to understand or, worse, incomprehensible.

An example following the standard includes FindLength(), and examples that do not follow the standard include abcdefg(), cccc(), and aa2020().

*E. Full Words*

This standard requires the use of full words as opposed to single-letter identifier names. This standard disallows names such as a, b, x1, x2 as method names. The reasoning for this standard is that there is a body of research specifically on the (mis)use of single-letter names. Hofmeister et al. [61] argue that meaningful, full word identifier names activate context semantics that allows developers to evaluate code against its purpose. Lawrie et al. [4] show that better comprehension occurs when full-word identifiers are used rather than single-letter identifiers. Lawrie et al. [62] assert that it is important for the identifier names to clearly communicate the concepts that they represent. Lawrie et al. [63] insist that informative identifiers are composed of full (natural language) words. Also, they imply identifier quality correlates to the use of dictionary words and coherent abbreviations.

An example that follows the standard is dbConnection(). An example that does not follow it is c() for the same method.

*F. Idioms and Slang*

The method name should not contain personal expressions, idioms, or slang. This standard is a special case of both the dictionary-term and full-word standards. We include it as these slang terms can be dictionary terms or full words but have no meaning in the context of the applications problem or solution domain of the application. In programming communities, using slang and idioms is sometimes referred to as being a *cute* practice [31]. Cuteness in code often appears in the form of colloquialisms or slang [31]. Programmers typically use it for humor and entertainment; however, Martin [31] insists on choosing clarity over entertainment value.

Examples not following the standard are personal names, `fido()`, idioms, `cutting_corners()`, and slang, `CurveBall()`.

*G. Abbreviations & Acronyms*

If a method name contains words that are abbreviations or acronyms, they should be well known or standard. Abbreviations or acronyms used in method names should be standard ones used by the organization or domain.

According to Lemire [64], he call unfamiliar abbreviations *evil abbreviations*, because some abbreviations are very hard to understand by programmers. In the book on rules and recommendations for programming in C++, Henricson [65] asserts that names should avoid abbreviations that are not generally accepted. Hofmeister et al. [61] claim that shorter identifier names take longer to comprehend. Schankin et al. [66] assert that descriptive compound identifier names improve source code comprehension. There is also research on expanding abbreviations in source code identifiers to mitigate such issues. Lawrie and Binkley [67], Alatawi et al. [68], Corazza [69], Fry [70], and Newman et al. [71, 72] discuss different types of abbreviations and study their distribution in various software artifacts; the source code, comments, software documentation, language documentation, and an English dictionary. They discuss the shortcomings and effectiveness of modern abbreviation expansion techniques as well as the most frequent locations of abbreviations and their expansions in the artifacts they use.

Acronyms used in method names should be standard ones used by the organization or domain. Hill et al. [73], Corazza et al. [69], and Newman et al. [71] discuss acronyms in source code and describe them as a name shorting made from the first letters of each word. Proper uses of acronyms include standards or protocol names such as URL or SQL.

Examples that follow the abbreviation standard are: `getStr()`, `pyConnection()`, `get_algo()`, `db_connection()` and contain abbreviations for string, python, algorithm and database respectively. Examples that do not follow the standard are: `repr()`, as the programmer cannot be sure about the correct expansion of this method name. Is it `repair()`? or `representation()`? Another example is `getProtoNameNode()`. The abbreviation `Proto` can stand for `Protocol` or `Prototype`. Examples that follow the acronym standard include `GUI_interface()`, `get_URL()`, `get_FIFO()`, and `DOM_tree()`. Examples that do not follow the standard are `get_QWE()` and `SendAAAA()`.

*H. Prefix/Suffix*

Method names should not contain a prefix or suffix that is a term from the system. These are sometimes referred to as preambles [12][59], which are a specific subset of prefixes. Since we are not limiting ourselves to the preamble subset, we use prefix/suffix. This standard does not apply to languages such as C that do not have namespaces. In languages without namespaces, it is common to add a prefix or suffix to differentiate subsystems. However, systems written in languages that do support namespaces should not use this idiom. In [31], Martin asserts that people quickly learn to ignore the prefix (or suffix) to see the meaningful part of the name and prefixes become unseen clutter and a marker of older code. Hungarian notations prefix identifier names with single letters that represent type, quantity, or scope. In this context, Martin [31] believes that names should not include type or scope information. He argues that prefixes such as *m_* or *f* are useless in today's development environments. That is also true to project and/or subsystem encodings such as `vis_` (for visual imaging system), which he believes distract and are redundant. Another downside to the use of prefixes or suffixes in identifier names is the difficulty for automated approaches to analyze and determine which terms should be considered prefixes; simple frequency count is not enough to identify all preambles (and, therefore, prefixes) [12].

Examples of names with prefixes or suffixes are `gimpItemGetPath()` and `swift_stdlib_u_char()` where *gimp* and *swift* are the names of the software in which the method appears.

*I. Length*

There was a long debate about the most preferred method length in programming. *"Can we hope to reach a kind of agreement about the ideal method length in OO software?"* [74]. We believe that there needs to be a fixed maximum number of words in a method name. While there are exceptions to this standard, it provides a guideline to developers for the maximum length of a name. Several researchers discuss the relationship of identifier length to the descriptiveness of a name. Schankin et al. [66] perform an experimental study to compare long and more detailed identifier names against short ones. They confirm that longer, descriptive identifiers have a positive impact on code comprehension. Knuth [75] observed that more descriptive identifiers are a clear indicator of code quality and comprehensibility. Liblit et al. [51] discuss several characteristics of the syntax and structure of identifier names. Among their findings, they discuss the length of identifiers; and find a high standard deviation between the lengths of identifiers in different systems, but that name length tends to sit between ~1 and ~5 words on a small sample of large systems. In refactoring, long method names are a type of code smell [1].

Here is an example from the open-source framework *Mockito* (site.mockito.org), written in Java, that contains 15 words:

`returnfalseifnosetterwasfoundandifreportnosetterfoundisfalse()`.

III. SURVEY DESIGN & METHODOLOGY

To design and deploy the survey, we followed Kitchenham's [76] guidelines for personal opinion surveys in software engineering. Thus, we start by identifying the high-level objectives for our investigation as follows: 1) A general consensus of what makes a good method name; 2) Developer's attitudes of each method naming standard based on years of programming and programming language background.

## A. Survey Design & Delivery

We use Qualtrics (www.qualtrics.com) in the design and delivery of the survey. Qualtrics is a tool for building, distributing, collecting, and analyzing participant responses.

The survey has three sections: 1) The introduction with a brief overview of the survey and an estimated completion time (10-15 minutes). 2) The ten survey questions (see Table I) related to each method naming standard. 3) Demographic questions that collect information about the participants.

TABLE I. METHOD NAMING STANDARD. EACH PART OF THE STANDARD AND THE ASSOCIATED SURVEY QUESTION.

| # | Standard Name | Survey Question |
|---|---|---|
| 1 | Naming Style | The method name should use a standard naming style such as camelCase or underscore_case; Camel case uses upper case letters for each word. Underscore case uses "_" to separate words. |
| 2 | Grammatical Structure | The method names with multiple words should be in a grammatically correct sentence structure. |
| 3 | Verb Phrase | The method name should always contain a verb(s) or verb phrase that refers to the behavior of the method. |
| 4 | Dictionary Terms | Developers should use only natural language dictionary words and/or familiar/domain-relevant terms. |
| 5 | Full Words | The method name should use full words rather than a single letter to clearly indicate the task of the method. |
| 6 | Idioms and Slang | The method name should not contain personal expressions, idioms, or unknown slang. |
| 7 | Abbreviations | The method name should contain only known or standard (i.e., recognized by others within the company) abbreviated terms. A poor abbreviation is one that has multiple possible expansions, interpretations or is not typically used within the system domain. |
| 8 | Acronyms | The method name should contain only known or standard (i.e., recognized by others in the company) acronyms. A poor acronym is one that has multiple possible expansions, interpretations or is not typically used within the system domain. |
| 9 | Prefix/Suffix | The method name should not contain a prefix/suffix that is a term from the system. This standard does not apply to languages such as C that do not have namespaces. |
| 10 | Length | The maximum number of words in a name should not be greater than (slider provided from 0-15) |

## B. Design of the Survey Questions

To investigate what professionals think about the standards, we took each standard and created an initial set of closed and open-ended survey questions. The aim of each question was to get feedback from the professional about the applicability of the standard and how much a participant agrees or disagrees. These questions contain examples of method names that both follow and do not follow the standard. They are also given a Likert scale of *strongly agree, agree, disagree,* and *strongly disagree*. Additionally, to gain more understanding about the participant's perceptions of each standard, each standard also has an open-ended text box asking for comments, thoughts, or opinions. The survey is anonymous as this increases response rates [77], and leads to more candid responses. Also, participants were allowed to skip the questions they do not want to answer.

In the development of the questions, we first performed a small pilot study to evaluate the survey questions. The study included five expert developers who also provided feedback. The feedback allowed us to fine-tune the wording of the questions and the examples for each standard.

## C. Demographic Questions

The survey included demographic about the participants. These questions ask how much they adhere to coding standards at their workplaces, what languages they are familiar with, and their years of programming experience. This information helps us to discover more insights into how professionals from different backgrounds perceive the standards.

TABLE II. SURVEY DEMOGRAPHIC QUESTIONS

| Question | Possible Responses |
|---|---|
| At your place of work, are there strict naming conventions in their coding standards? | Very Strict, Strict, No Standards, Slightly Strict, Not Strict. |
| Please give a short description of the naming coding standards you use. | Text box |
| Which programming language are you familiar with? | C++, Python, Java, C, C#, JavaScript, other language. |
| How many years of programming experience do you have? | Less than a year, 1 - 2 years, 3 - 4 years, 5 - 9 years, 10+ years, prefer not to say. |

## D. Recruiting Participants

After finalizing the survey questions, we used multiple means to recruit survey participants. As feedback from software professionals is needed, we started with a systematic sampling approach [30] in which we sent personalized email invitations to software professionals working in large software companies such as Google, IBM, Intel, and Microsoft. Next, we individually invited software engineering researchers to participate, stating the need for expert developers. The invitation also includes using the snowball sampling technique in which the invitation asked the professional to forward the invitation to other professionals. Finally, we reached out to Stack Overflow co-founder (i.e., Jeff Atwood) as he is interested in coding standards and has a large social presence in the developer community. This step is non-systematic [30]; however, it resulted in the sharing of the survey with the Stack Overflow community.

## IV. DATA COLLECTION AND ANALYSIS

As stated previously, we use Qualtrics to deliver the survey and collect the responses. We also use it for analyzing and manipulating the quantitative part of the survey, i.e., closed questions.

For the open-ended questions (i.e., qualitative analysis), we use MAXQDA (maxqda.com) for thematic coding. We adopt an inductive approach for analyzing participants' attitudes in which themes are emergent while reading and going through participants' comments. Thematic analysis is a common mechanism for identifying, analyzing, and reporting patterns and themes within data [78].

We did intensive qualitative analysis on the comments to determine the different perspectives on the method standards. Fig. 2 shows the steps we take for analyzing the participants' comments. First, the comments are organized based on years of experience and programming language knowledge. Then the comments are scanned over to understand general trends. After gaining some understanding of the comments, they are assigned labels (e.g., camel case, readability, testing) of general topics mentioned. After this process is complete, the list of all labels is compiled. This list is used to create a set of themes (categories). Related comments are grouped based on topics (labels). The process also includes negotiating with other researchers about emerging themes and fine-tuning the topics as necessary.

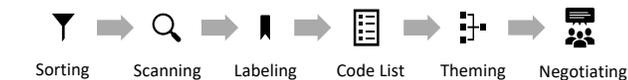

Sorting  Scanning  Labeling  Code List  Theming  Negotiating

Fig. 2. Steps for analyzing participants' comments on each standard

We found that participants often use metaphoric words for expressing their attitudes; those are also included in our analysis. Also, during the analysis, analytic memos are used as personal notes for denoting ideas, interpretations, or unfamiliar comments.

Fig. 3 gives a screenshot of the tool with each comment on the right and the various labels on the left. It allows the user to organize and categorize the comments in a visual manner. Counts of each label and theme are calculated. These counts are used to determine the most common topics participants mention, and some comments for these topics are provided in the next section.

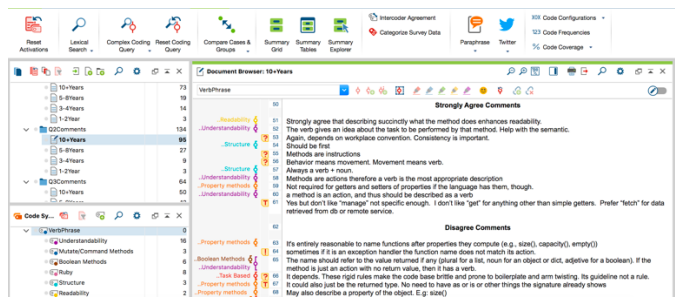

Fig. 3. Participant' comments analysis

TABLE III. PROGRAMMING LANGUAGE USED BY 1164 PARTICIPANTS.

| Language | Number of Responses | Percentage of Participants Knowing the Language |
|---|---|---|
| C++ | 472 | 40.6% |
| C | 484 | 41.7% |
| C# | 589 | 50.7% |
| Python | 635 | 54.7% |
| Java | 697 | 50.0% |
| JavaScript | 838 | 72.1% |
| Other Language | 479 | 41.2% |
| **Total responses** | **4194** | |
| **Avg # of Languages** | **4** | |

## V. RESULTS OF SURVEY

We received a total of 1604 responses to the final survey. This large amount of responses appears to demonstrate that the developer community is interested in this topic. We filtered out incomplete responses and only considered answers for practitioners with three years of experience and more. We did this to exclude responses from non-professionals (students). Additionally, there were only a small number (31) of responses in this category. Thus, the total responses considered in this work is 1162. The survey was available from March 13, 2020, until June 2, 2020. With the majority of responses coming in April. The average mean time to complete the survey is 15.94 minutes, close to our goal of 15 minutes. Fig. 4 shows the participant counts based on years of experience.

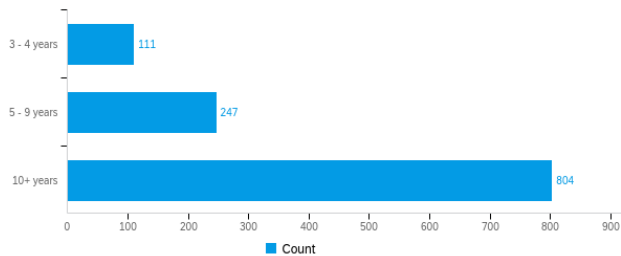

Fig. 4. Total participants according to years of experience

The majority, 70% (804), of the survey participants have more than 10 years of programming experience. 21% (247) have five to nine years of experience, and about 9% (111) have three to four years. Hence, we definitely met our objective to survey experienced developers.

A survey question instructed the participants to indicate all programming languages with which they have experience. Table III shows the programming language background of all the participants. There is a nice diversity in language background with between 40% and 50% of the participants knowing C++, C, Java, C#, and Python. A slightly larger group (72%) of the participants know JavaScript. Around 40% of participants responded that they have experience with other languages not on the list. Of these, PHP, Ruby, and Go are each known by around 100 participants. Another 35 languages are mentioned, but none by more than 50 participants (most by less than 15). On average, participants are familiar with 4 programming languages.

The responses to the ten survey questions on each standard appear in Table IV. Overall, the vast majority of the responses are in agreement (i.e., Strongly Agree or Agree) for all of the standards. All are at 80% or more agreement with many at 90% or more. Only two standards (i.e., Verb Phrase and Grammatical Structure) have a notable disagreement. Fig. 5 illustrates the participants' agreement level in a diverging stacked bar chart. As the answers to the length question are not on a Likert scale, the results for this standard appear in a following subsection.

We now discuss the results for each of the standards separately. The results are also analyzed based on years of experience and programming language knowledge. Additionally, we give a qualitative assessment of participant responses for each standard.

TABLE IV. SURVEY RESULTS

| Question | Response Type | | | |
|---|---|---|---|---|
| | *Strongly Agree* | *Agree* | *Disagree* | *Strongly Disagree* |
| **Naming Style** | 87.20% | 11.51% | 0.87% | 0.43% |
| **Grammatical Structure** | 35.21% | 43.51% | 19.29% | 1.99% |
| **Verb Phrase** | 41.50% | 43.14% | 12.68% | 2.67% |
| **Dictionary Terms** | 73.89% | 23.33% | 2.52% | 0.26% |
| **Full Words** | 74.20% | 22.69% | 2.85% | 0.26% |
| **Abbreviations** | 58.33% | 35.16% | 5.30% | 1.22% |
| **Acronyms** | 66.03% | 29.79% | 3.83% | 0.35% |
| **Idioms and Slang** | 51.99% | 37.09% | 9.79% | 1.13% |
| **Prefix/Suffix** | 49.34% | 39.98% | 9.89% | 0.79% |

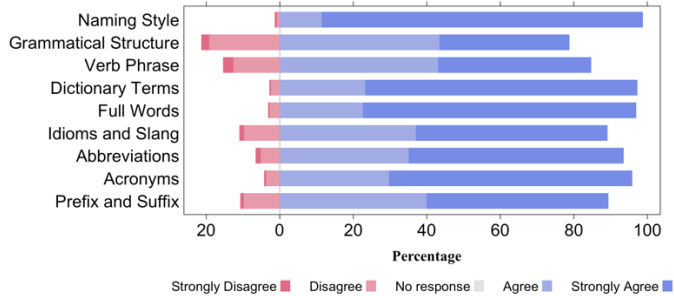

Fig. 5. Participants answers in a diverging stacked bar chart

### A. Naming Style

This question aims to investigate participants' opinions about using a naming style such as camelCase or underscore in naming methods. The participants are in very much in agreement (99%) on using a consistent naming style for methods (TABLE IV. ). They support their agreement with the fact that a naming style promotes readability. There are 105 comments for this standard. Developers insist that using a consistent naming style is a key factor when naming methods across a project. Additionally, developers believe that using a specific naming style is programming-language dependent, e.g., PascalCase for C# methods. Those who disagree with this standard argue about their preferences towards using a specific naming style. Otherwise, they agree with the standard.

> *"The naming style should follow language recommended naming style." (Participant with 5-9 years of experience)*

Several participants mention cases where they prefer using a particular naming style over another. For example, some participants state that they like using the underscore naming style with constants, and some mention that they often see underscores in test classes containing multiple tests, but not method names.

> *"I dislike underscores in method names. Save them for constants" (Participant with 10+ years of experience)*

### B. Grammatical Structure

The question for this standard checks professionals' attitudes towards having a grammatically correct sentence structure in method names (that contain multiple words). Seventy-nine percent of developers agree that method names containing multiple words should be in a grammatically correct sentence structure. Participants who strongly agree on this standard support their opinion with their belief that grammar structures motivate readability and comprehension. There are 93 comments concerning this standard. Participants insist that clarity is the key to constructing method names.

Others generally prefer grammatically correct structures, but they may break it if the name does not deliver the exact intent of the task. Some also believe that grouping by a feature or an entity instead of verbs make more sense in some cases. Additionally, participants against this standard argue that not everybody knows English, so flexibility in this standard is needed.

> *"Generally true, except where unusual syntax clears up ambiguity." (Participant with 10+ years of experience)*

### C. Verb Phrase

The question for this standard is aimed at exploring developers' views about requiring a verb(s) or a verb phrase in the method name, which refers to the behavior of the method. The participants have an agreement of 85% concerning this standard. They believe that adopting such a naming practice enhances understandability.

> *"The verb gives an idea about the task to be performed by that method. Help with the semantic." (Participant with 10+ years of experience)*

When we examined the participants' agreement according to their years of experience, we find that among programmers with 10 or more years of experience, 83% agreed, and 16% disagreed. Similarly, for the other two experience groups, there is a 12-14% disagreement. There are 122 comments for this question.

A number of the comments mention a specific naming case: how accessor and mutator methods (e.g., getter, setter, property, predicate) are named. The question is if they need to be a verb phrase as many documented standards encourage. On this regard, Tan [36] feels naming predicates by adding the prefix check or get is one appropriate way to name Boolean functions. Piater [37] also share the same view and believe that appending is or has to predicates is more suitable.

However, several comments from developers who disagree with this form and feel that adding the verb get_, set_, or is_ is no longer necessary or useful. Thus, they advocate for instead of getLength() that length() is better, likewise isEmpty() should be just empty(). As programming languages evolve and change, so do naming standards. We see this a such a case. The naming standards are evolving along with how developers interpret different types of methods. When object-oriented programming started to become widely used in the 1990s, it was the practice to use such leading verbs (some languages automatically generate these types of accessors/mutators with such names). Clearly, there is a change in perception of how these should now be named. This perception change most likely reflects the ubiquitous nature of the accessor/mutator concept.

> *"I prefer naming a getter without the "get" prefix." (Participant with 10+ years of experience)*

Some participants also mention that the inclusion of verb phrases depends on the method goal, e.g., command, query, predicts, etc. Others mention that it is more useful to name the method after what it returns, not necessarily including an action verb, so they feel including a verb in the method name is a task-based practice.

*D. Dictionary Terms*

This question seeks developers' opinions about using only natural language dictionary words and/or familiar/domain-relevant terms while naming method names. Dictionary terms are the clearest way to communicate, and non-dictionary words negatively impact developers' comprehension. Results on this question show that 97% of participants agree on this standard. Participants agree that using dictionary terms supports the meaningfulness of a method name as well as readability and comprehension. We received 71 comments concerning this standard. The majority of comments assure that words chosen in method names should always be meaningful and descriptive, and this applies to the project-specific terms as well.

> *"Not just a known word - it should be a perfect word(s) both by meaning - I can spend even hours searching through dictionary and synonym lists for the perfect name." (Participant with 5-9 years of experience)*

A theme that emerged among participant opinions on using dictionary terms is to consider common abbreviations and project-specific terms. Participants who disagree with this standard only argue about some common abbreviations that can enhance the searchability across the code base, otherwise they agree with the standard.

*E. Full Words*

This question aims at professionals' attitudes towards using full words and no single letters in method names. Participants are in 97% agreement that the name should contain full words to clearly indicate the task of the method rather than a single letter. They believe that having full words is a key to supporting readability, and using single letters in a method name leads to unreadable code. Developers provided a total of 47 comments on this question. They believe that single letter method names do not give a clear indication even for the actual programmer when he gets back to code after a long time.

> *"Single letters can easily be misinterpreted." (Participant with 10+ years of experience)*

*F. Abbreviations & Acronyms*

Both questions on abbreviation and acronyms target developers' beliefs about using unfamiliar shortenings. The participants are in 94% of agreement that the method name should contain only known or standard abbreviated terms. 96% of the survey participants were in agreement that the method name should only contain known or standard acronyms as long as they are related to the project and can be interpreted from the context or explained in a comment. Their rationale is to always think about other later programmers or newcomers who do not know much about the project. A total of 109 comments on abbreviations, 53 comments on acronyms are received.

> *"Too open for misunderstanding, abbreviation meaning within business/domain change over time". (Participant with 10+ years of experience)*

A respondent mention that the auto-complete feature in editors made this easy, and there is no need for abbreviations.

> *"No need to abbreviate, all editors have autocomplete." (Participant with 10+ years of experience)*

Disagreeing responses only insist that the context could play some role in applying these two standards, otherwise they agree.

> *"Sometimes there are context-local acronyms that make sense to exist in a team's codebase" so there is no issue using these acronyms." (Participant with 10+ years of experience)*

*G. Idioms and Slang*

This question concerns developers' beliefs about using idioms and slang in method names. 89% of participants agree that the method name should not contain personal expressions, idioms, or unknown slang. Their main concern is to consider fellow programmers' as the use of idiom and slang can negatively impact their comprehension and understandability of the code. A total of 71 comments were on idioms and slang. A few participants who disagree argue that it is a personal preference practice; unless it is a team project.

> *"This is personal preference unless the team is distributed and consists of diverse cultural backgrounds." (Participant with 10+ years of experience)*

A few comments argue on using unfamiliar and clever names if it fits the task very well.

> *"I only slightly disagree, especially in cases where something clever fits very well". (Participant with 5-9 years of experience)*

In this regard, Martin [31], asserts that if names are too clever, they are only memorable to people who share the author's sense of humor, and only as long as these people remember the joke.

*H. Prefix/Suffix*

This question addresses developers' opinions about using a prefix or suffix in method names. 89% of participants agreed that a method name should not contain a prefix or suffix that is a term from the system. They support their agreement with the notion of not being repetitive over source code. There is a total of 62 comments on this standard. Participants generally believe that not adding a class or namespace name is sensible except for some cases in which the developer wants to differentiate between different packages embeddings as prefix and suffix helps in identifying the component the developer is referring to.

> *"In some cases it helps to identify the component the developer is referring to." (Participant with 10+ years of experience)*

*I. Length*

The purpose of this question is to find out professionals' preferences on the maximum length of a method name. The distribution of the responses for this standard is in Fig. 6. The median and mode are both 5 words. 81% of the responses are 5 ± 2 words. Hence there is a strong agreement among the participants that method names should be relativity short and descriptive and between 3 and 7 words in length. This agreement aligns with research on human short-term memory.

The work of Miller [79] on chunking gives the typical maximum of human short term memory at $7 \pm 2$ chunks, where a chunk is an easy to remember aggregate, such as a word.

> *"Method names should be short but descriptive. They should convey what it does and nothing more."* (Participant with 10+ years of experience)

Beyond 8 words, there are few responses until the maximum number allowed as a response in the survey question. Also, there are almost no responses below 3 words. There are 159 comments on this standard. Upon inspecting these comments, we found that people who choose 15 words as the maximum length argue that test drivers often need to have long names. That is, the method that runs a test (or set of tests) of the application needs to fully describe the test scenario/case.

> *"Usually ok for tests to break this rule and have long names that explain what the test does."* (Participant with 10+ years of experience)

These comments point to a separate naming standard that deals specifically with test cases. Tests are not a part of a software application. That is, they are not part of an observable feature of the system. Rather they are part of the build management system of the application. Additionally, various unit testing frameworks have their own naming conventions for each test case. Hence, we feel that a separate naming standard for methods that implement testing is clearly warranted and should be articulated for a project (or organization). The standard for such methods will be different from methods that implement the functionality of the application and may need to be quite lengthy.

> *"Test methods have more rules such that they should contain the thing under test, the key condition being set up, and the expected result."* (participant with 10+ years of experience)

Given the survey results, it appears that developers do not want method names to ever exceed 8 words, and more typically should be no longer than 5 words.

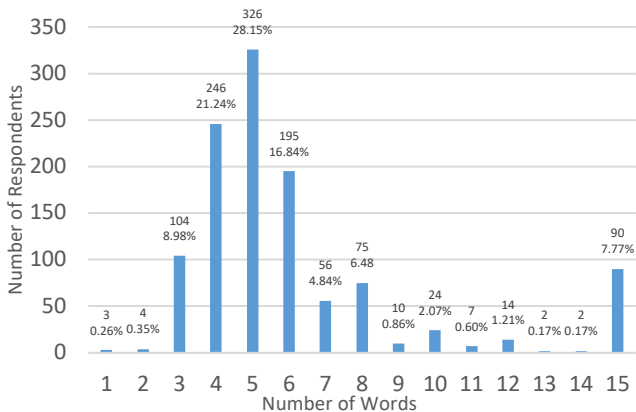

Fig. 6. Maximum method length preferred by professionals

### J. Additional Developer Feedback

In order to cover all the method naming aspects, we asked the professionals to contribute any other standards they felt important. We received 247 comments on this topic. After handling the analysis procedure, we found some interesting emerging themes that are now discuss.

A number of respondents indicate the importance of considering the scope of methods while naming, i.e., public methods and private methods, and, more specifically, when choosing a naming style.

> *"Differentiation between public and private with lower uppercase or underscore."* (Participant with 10+ years of experience)

Another interesting opinion suggests that developers should not repeat the argument names in the method name, e.g., `saveFirstName(firstName)`. Some comments also assert that programmers should not use Hungarian notation while naming methods.

> *"you should never use Hungarian notation, like `boolIsValid()` or `stringGetName()`."* (Participant with 10+ years of experience)

Other comments suggest that programmers should not use the return type in the method name.

> *"no type hint in method name (example: `make_string_hello_world` vs. `make_hello_world`)"* (Participant with 10+ years of experience)

Many comments insist that method names, when used in a class, should not be prefixed by the class name. They also recommend avoiding methods with names that can be confused with each other (e.g., `find()` and `search()` in the same class).

## VI. DISCUSSION

Let us now address each research question in the context of the findings and survey results.

### A. RQ1

To what extent do software professionals support the method naming standards conveyed in the survey?

Based on Fig. 5, we see that there is wide support of the presented method naming standard. The only standards with substantial disagreement deal with the grammatical structure and verb phrase. Much of the disagreement has to do with special cases (e.g., accessor/mutators). Even then, the agreement on these two standards is near 80%.

Given the diversity of developer background in terms of programming language knowledge, it also appears that this result generalizes to most any programming language.

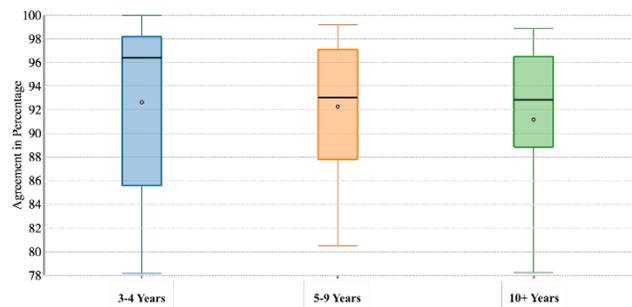

Fig. 7. Participants agreement across all the standards by experience level

### B. RQ2

Do years of programming experience impact how professionals respond to the survey questions?

We examine the responses to determine if years of programming experience impact how professionals respond to the survey questions. We found that years of experience does not have any impact on the response to the survey questions concerning each standard. Fig. 7 shows a box plot on the agreement percentages across all the standards by experience level. It is noticeable that the mean and the median of all the group participants are consistent.

We ran a chi-square test on each individual question and found no significant difference in all but one of the questions. There is a significant difference (0.04 p-value) in how more experience developers responded to the abbreviation standard. They are more comfortable with using abbreviations and disagreed with the standard a bit more than less experienced developers. We hypothesize that, as developers gain experience, they become familiar with an increasing set of abbreviations and grow to prefer these abbreviations over their expansions. The question of when an abbreviation should be expanded is open [12] but this result implies that developer experience should be taken into account in future research on the topic.

### C. RQ3

Does the programming language used impact how professionals respond to the survey questions?

Fig. 8 gives a boxplot of the agreement percentages for each question across programming language experience. This shows that there are very few outliers from the standard deviation for each question. That is, everyone answers in a very similar manner no matter their language experience. There are outliers for three standards. Developers with Java and C# experience are outliers (small amount) for naming style. C programmers are outliers (small amount) for dictionary terms. Lastly, there is an outlier for verb phrases, which is from the other languages' category.

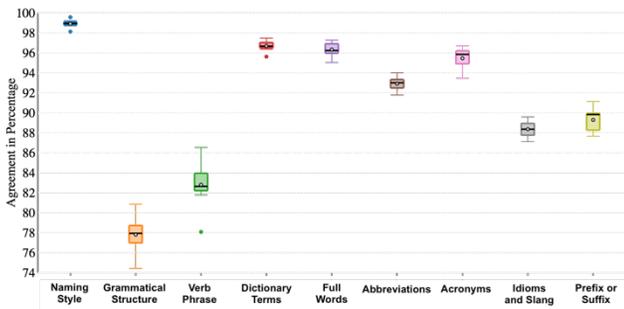

Fig. 8. Agreement percentages across all the standards by programming language experience

### D. RQ4

What are professionals' perceptions of each part of the method naming standard in the survey?

Professionals generally seem to accept the standards based on the comments and responses. We received at least 45 comments for each of the standards in the survey. These comments went through an in-depth qualitative analysis to draw our final conclusions. Upon finishing the survey, developers mention that they learned new method naming guidelines and welcomed such guidelines as they believe that is a positive step towards clarity and simplicity. They insist that consistency is a key factor in method naming.

The survey also includes a demographic question asking how strict naming conventions are in the participant's workplace. This question is to determine the general prevalence of naming standards in organizations and how strict they are adhered. The results show that approximately 60% of the participants either do not have strict naming standards or do not have any standard at all. Based on the provided comments, we found that a number of developers rely on code reviews to keep track of their naming and flag poor and inconsistent naming choices. There were no particular preferences or barriers concerning the standards except the fact that some standards need a little flexibility, as we have discussed.

### VII. THREATS TO VALIDITY

As our primary instrument is a survey, we carefully formulated the questions considering the guidelines provided in [80]. To address construct validity [81], we performed a set of discussions with a focus group of experts in the field to determine the quality of the questions and the survey objectives. Several question drafts were examined and fine-tuned before publishing the final survey to ensure that the questions were clear. The standards in the survey are obtained from the software engineering literature and published coding standards. We address the internal consistency of the survey by only taking into consideration questions that examine method naming quality. Internal validity is another concern we addressed by considering experts evaluation for each provided question and taking into account any feedback that helps in achieving the main objective of the study. With regards to external validity, we believe that the quantitative part of the survey is generalizable, in which we believe that if the same questions were given to another similar population, the outcomes would be the same or similar. Group discussion is also used during the qualitative analysis for the comments to support the credibility of the results.

### VIII. CONCLUSIONS

The results of a survey to assess professional developer's opinions on method naming standards is presented. Overall, developers are very much in agreement with the given standards. From the written responses we gleaned that developers are supportive of clearly articulating method naming standards and feel it has a positive impact to code comprehension.

The results of the survey provide valuable implications. The first is that the method naming standard is valid or at least widely acceptable by developers. The second is that the standard is complete (with the addition of additional guidelines for access/mutators and test drivers). This is important in that we can use these standards to construct automated tools that assess the quality of method names with a high degree of accuracy. With the use of natural language processing tools and dictionaries all aspects of the standard can be implemented in such a tool to a large degree. Such a tool would be invaluable for developers and during code reviews. This could also lead to recommendation system for method names.

ACKNOWLEDGMENT

We thank all the participants for their responses. We also thank all who forwarded our request to others in their organization.